\documentstyle[aps,epsfig,multicol]{revtex}
\newcommand{\be}{\begin{equation}}
\newcommand{\ee}{\end{equation}}
\newcommand{\bea}{\begin{eqnarray}}
\newcommand{\eea}{\end{eqnarray}}
\newcommand{\HH}{{\cal H}}

\newcommand{\p}{\partial}
\newcommand{\s}{\sigma}
\newcommand{\la}{\langle}
\newcommand{\ra}{\rangle}

\newcommand{\lp}{\left(}
\newcommand{\rp}{\right)}
\renewcommand{\vec}[1]{{\bf #1}}

\begin{document}
\title{Solitons and Rabi Oscillations in a Time-Dependent BCS Pairing Problem}
\author{R.\,A. Barankov$^1$, L.\,S. Levitov$^1$ and
B.\,Z. Spivak$^2$}

\address{$^1$Department of Physics, Massachusetts Institute of Technology,
77 Massachusetts Ave, Cambridge, MA 02139}
\address{$^2$Department of Physics, University of Washington,
Seattle, WA 98195}

\maketitle
\begin{abstract}
Motivated by recent efforts to achieve cold fermions pairing near
a Feshbach resonance, we consider the dynamics 
of formation of the
Bardeen-Cooper-Schrieffer (BCS) state.
At times shorter than
the quasiparticle energy relaxation time, 
after the interaction is turned on,
the dynamics of the system
is nondissipative.
We show that this collective nonlinear evolution of the
BCS-Bogoliubov amplitudes $u_p$, $v_p$, along with
the pairing function $\Delta$, is an integrable dynamical problem,
and obtain a family of exact solutions
in the form of single solitons and soliton trains.
We interpret the collective oscillations
as Bloch precession of Anderson pseudospins,
where each soliton causes a pseudospin $2\pi$ Rabi rotation.
Numerical simulations demonstrate robustness 
of the solitons with respect to noise and damping.
\end{abstract}


\begin{multicols}{2}

\narrowtext
Dilute fermionic alkali gases cooled below degeneracy~\cite{fermion_expts}
are expected to host the paired BCS state~\cite{Stoof96,Ho99}.
One of the unique and attractive features of this system which makes it
qualitatively different from superconducting metals is
the possibility to control the strength of pairing interaction and change it 
by using the Feshbach
resonances~\cite{Bohn00,Holland01,Kokkelmans02,Milstein02,Ohashi,Stajic03}.
Pairing enhancement near the resonances, as well as the high coherence of atomic systems,
can allow to realize the strong coupling 
BCS regime, as well as to explore the time dynamics of the paired
state. 
Since the characteristic energy scales in atomic vapors are relatively low,
one can perform time-resolved measurements on the intrinsic microscopic time scales,
and explore a range of fundamentally new phenomena in the time dynamics 
of the paired state. These new prospects helped to revive interest in some of 
the basic issues of the BCS pairing 
problem~\cite{Heiselberg00,Baranov02,Heiselberg02,Search021,Carlson03,Heiselberg03}.

In particular, one of the important questions has to do with 
the dynamics and characteristic times of 
formation of the BCS state in cold gases~\cite{Houbiers99}.
The dynamics of the superconducting state
in metals, described by BCS theory, has been a subject of 
active research for a long time~\cite{Tinkham}.
Generally, there are two important time scales to be considered: 
the quasiparticle energy relaxation
time $\tau_{\epsilon}$, and the characteristic time of the order parameter 
change, $\tau_{\Delta}$, estimated as the inverse increment of Cooper 
instability~\cite{Schmid66,Abrahams66}.
For $\tau_{\epsilon}\ll \tau_{\Delta}$, quasiparticles
quickly reach local equilibrium 
parameterized by a time-dependent order parameter $\Delta(t)$. 
In this case the dynamics is described by the time-dependent Ginzburg-Landau
(TDGL) equation for $\Delta(t)$, 
with the relaxation time scale {\it ca.} $\tau_{\Delta}$. 
However, as noted by Gorkov and Eliashberg~\cite{Gorkov68},
this scenario applies only to relatively exotic situations,
including a close proximity of a transition point,
$\Delta^{2}/T_{c}\ll \tau^{-1}_{\epsilon}$, 
or a fast pair breaking 
(e.g., due to paramagnetic impurities).

In the opposite limit, which takes place at the temperatures not too close
to critical,
\be\label{eq:tau_epsilon>tau_delta}
\tau_{\epsilon}\gg \tau_{\Delta}
\ee
the system dynamics is usually described by
Boltzmann kinetic equation for quasiparticles
and a self-consistent equation for $\Delta(t)$ connecting it
with the quasiparticle distribution function
locally in time~\cite{Aronov74,LarkinOvchinikov}. The validity of this approach requires, in
addition to (\ref{eq:tau_epsilon>tau_delta}), that
the variation in time of both the
quasiparticle distribution and the external parameters 
is sufficiently slow on the $\tau_{\Delta}$ time scale.
The origin of this criterion is 
seen most easily if one notes that
the order parameter can exhibit free oscillations with a frequency 
{\it ca.} $\tau_{\Delta}^{-1}$~\cite{Anderson58} (also, see below). 
Thus, at a slow parameter variation
the order parameter $\Delta(t)$
and the quasiparticle spectrum 
will adiabatically follow
the changes of the quasiparticle distribution, without exciting
the oscillations of $\Delta$.  This approximation is
relevant to the majority of situations in superconducting metals
because of the relatively big value of $\Delta$ 
and the difficulty of changing the external parameters and making
measurements on the time scale of $\tau_{\Delta}$.

From this point of view, the cold fermionic gases present a completely 
new situation. The energy relaxation in these systems is quite slow,
while the external parameters, such as the detuning from Feshbach resonance,
can change very quickly on the time scale of $\tau_{\Delta}$. Thus 
the BCS correlations
in this case build up in a coherent fashion while the system is out of thermal
equilibrium. In such a situation, theory must account not only for the order
parameter evolution, but also for the full dynamics of individual Cooper pairs
and quasiparticles. It seems to be desirable
to understand better this 
`fast BCS buildup' regime, since the lifetime of the gas samples is finite, 
while the time-resolved measurements can easily be performed with 
resolution better than $\tau_{\Delta}$.

In this article we consider the situation when the pairing interaction is switched
on essentially instantly, on a time scale $\tau_0\ll \tau_{\Delta},\tau_{\epsilon}$.  
We shall discuss only the spatially uniform situation,
relevant for samples of finite size, and explore
the time evolution of the unpaired ideal Fermi gas,
which is unstable with respect to Cooper pairing. 
At not too long times, $t\ll \tau_{\epsilon}$, the dynamics of the system
is governed by nondissipative equations which conserve both the entropy and the
total energy. 
A stationary superconducting state 
with the same energy as that of the initial state
would have more quasiparticles, and thus would have
entropy {\it higher} than the initial Fermi gas entropy.
This argument suggests that
the system can reach a stationary (not necessarily
equilibrium) state only at a very long time of the order
$\tau_{\epsilon}$. On shorter times, which nevertheless
can exceed $\tau_{\Delta}$), the system will 
exhibit a nonlinear time evolution. During this period of time
{\it the concept of the quasiparticle spectrum is irrelevant} 
and theory can rely neither on the kinetic equation, nor on TDGL equation.
Below, we present an approach which
describes the BCS state buildup and accounts for coherent dynamics 
of individual Cooper pairs.
We shall focus on the zero temperature case, 
when $\tau_{\Delta}\simeq \Delta^{-1}$, and
show that the result of Cooper instability
is a periodic oscillation $\Delta(t)$ 
having the form of a soliton train.

This regime can be described by the BCS hamiltonian
\be\label{eq:Hbcs}
\HH = \sum_{\vec p,\,\s} 
\epsilon_{\vec p} a^+_{\vec p,\s}a_{\vec p,\s}
-\frac{\lambda(t)}2 
\sum_{\vec p,\vec q} a^+_{\vec p\,\uparrow}a^+_{-\vec p\,\downarrow}a_{-\vec q\,\downarrow}a_{\vec q\,\uparrow}
\,,
\ee
with the coupling turned on abruptly, $\lambda(t)=\lambda\theta(t-t_\ast)$.

The main result of this work is that the time-dependent problem
(\ref{eq:Hbcs})
is integrable.
We generalize the BCS solution, which is exact for the separable 
pairing Hamiltonian (\ref{eq:Hbcs}), and demonstrate that
at $t>t_\ast$
the many-body state
evolves as
\be\label{eq:Psi_bcs}
|\Psi(t)\ra = \prod_{\vec p}\lp u_{\vec p}(t)+v_{\vec p}(t)a^+_{\vec p,\uparrow}a^+_{-\vec p,\downarrow}\rp |0\ra
\ee
The Bogoliubov mean field treatment,
which gives a state of the form (\ref{eq:Psi_bcs}),
relies on the `infinite range' form of
the pairing interaction in (\ref{eq:Hbcs})
(i.e., equal coupling strength of all  
$(\vec p,-\vec p)$, $(\vec q,-\vec q)$).  
Since the latter does not depend
on the system being in equilibrium,
one can introduce a time-dependent mean field
pairing function
\be\label{eq:Delta_BCS}
\Delta(t)=\lambda\sum_{\vec p}u_{\vec p}(t)v_{\vec p}^\ast(t)
\ee
The amplitudes $u_{\vec p}(t)$, $v_{\vec p}(t)$
can be obtained from the Bogoliubov-deGennes equation
\be\label{eq:Bogoliubov_deGennes}
i\p_t \lp\matrix{ u_{\vec p} \cr v_{\vec p}}\rp =
\lp\matrix{ \epsilon_{\vec p} & \Delta \cr \Delta^\ast & -\epsilon_{\vec p}}\rp
\lp\matrix{ u_{\vec p} \cr v_{\vec p}}\rp
\ee
to be solved along with the selfconsistency
condition (\ref{eq:Delta_BCS}).

We recall that the unpaired state is a selfconsistent, albeit unstable, solution of 
Eqs.\,(\ref{eq:Bogoliubov_deGennes}),(\ref{eq:Delta_BCS})
with $\Delta=0$, $T=0$:
$
u^{(0)}_{\vec p}(t) = e^{-i\epsilon_{\vec p}t}\theta(\epsilon_{\vec p})
$, $
v^{(0)}_{\vec p}(t) = e^{i\epsilon_{\vec p}t}\theta(-\epsilon_{\vec p})
$,
The stability analysis~\cite{Abrahams66}
shows that
the deviation from the unpaired state grows as $\Delta(t)
\propto e^{\gamma t}e^{-i\omega t}$,
\be\label{eq:exp(gamma_t)}
\delta u_{\vec p}(t)=
\frac{\Delta(t) v^{(0)}_{\vec p}(t)}{i\gamma - 2\epsilon_{\vec p}+\omega}
,\quad
\delta v_{\vec p}(t)=
\frac{\Delta^\ast(t) u^{(0)}_{\vec p}(t)}{i\gamma + 2\epsilon_{\vec p}-\omega}
\ee
with the growth exponent $\gamma$ and the constant $\omega$ given by
\be\label{eq:instability_exponent}
1=\lambda\sum_{\vec p}\frac{{\rm sgn}\,\epsilon_{\vec p}}{2\epsilon_{\vec p}-\zeta}
,\quad
\zeta=\omega+i\gamma
\ee
The electron-hole symmetry near the Fermi level,
$N(\epsilon_{\vec p})=N(-\epsilon_{\vec p})$, makes
the frequency shift $\omega$ vanish.
Using the similarity between Eq.\,(\ref{eq:instability_exponent})
and the BCS gap equation at $T=0$,
the exponent $\gamma$ can be shown to be
close to the BCS gap value
$\Delta_0$. (In fact, in the constant density
of states approximation, the two quantities coincide:
$\gamma=\Delta_0$.)
Thus, we estimate the time $\tau_{\Delta}\simeq \Delta_0^{-1}$.

The interaction switching, while nonadiabatic, must also be
gentle enough not to overheat the fermions 
above $T\simeq\Delta_0$. 
The effective temperature $T_{\rm eff}$ after switching
can be estimated from the total energy increase, giving
\be\label{eq:Teff}
a T_{\rm eff}^2=\!\!\!
\sum_{\omega,\,1...4}
\!\!\!\hbar\omega
n_1n_2(1-n_3)(1-n_4)|\lambda_\omega|^2
\delta(\hbar\omega-\delta E)
\ee
with $n_i=n(\epsilon_{\vec p_i})$, 
$\delta E=\epsilon_{\vec p_3}+\epsilon_{\vec p_4}-\epsilon_{\vec p_1}-\epsilon_{\vec p_2}$, $a=\frac{\pi^2}6\nu$. 
For the interaction switching from $0$ to $\lambda$ 
over a characteristic time $\tau_0$,
the RHS of (\ref{eq:Teff}) is of the order
of $\lambda^2\tau_0^{-3}$. Thus the cold $T_{\rm eff}$
condition is
$E_F\tau_0\gg (\lambda n/\Delta_0)^{2/3}$, which is compatible with the
nonadiabaticity requirement $\tau_0\ll \tau_{\Delta}$.

At $T=0$, a soliton solution of
Eqs.\,(\ref{eq:Bogoliubov_deGennes}),(\ref{eq:Delta_BCS})
can be constructed most
naturally in terms of the variable
\be
w_{\vec p}=\cases{u_{\vec p}/v_{\vec p}, & $\epsilon_{\vec p}>0$\cr
v_{\vec p}/u_{\vec p}, & $\epsilon_{\vec p}<0$}
\ee
Consider first the case $\epsilon_{\vec p}>0$.
From Eq.\,(\ref{eq:Bogoliubov_deGennes}) we obtain
\be\label{eq:w_p}
i\p_t w_{\vec p}
= 2\epsilon_{\vec p} w_{\vec p} + \Delta(t) - \Delta^\ast(t) w_{\vec p}^2
\ee
where $\Delta(t)$ is a function of time determined from (\ref{eq:Delta_BCS}).

\begin{figure}[t]
\centerline{
\begin{minipage}[t]{1.2in}
\vspace{0.0pt}
\centering
\includegraphics[width=1.2in]{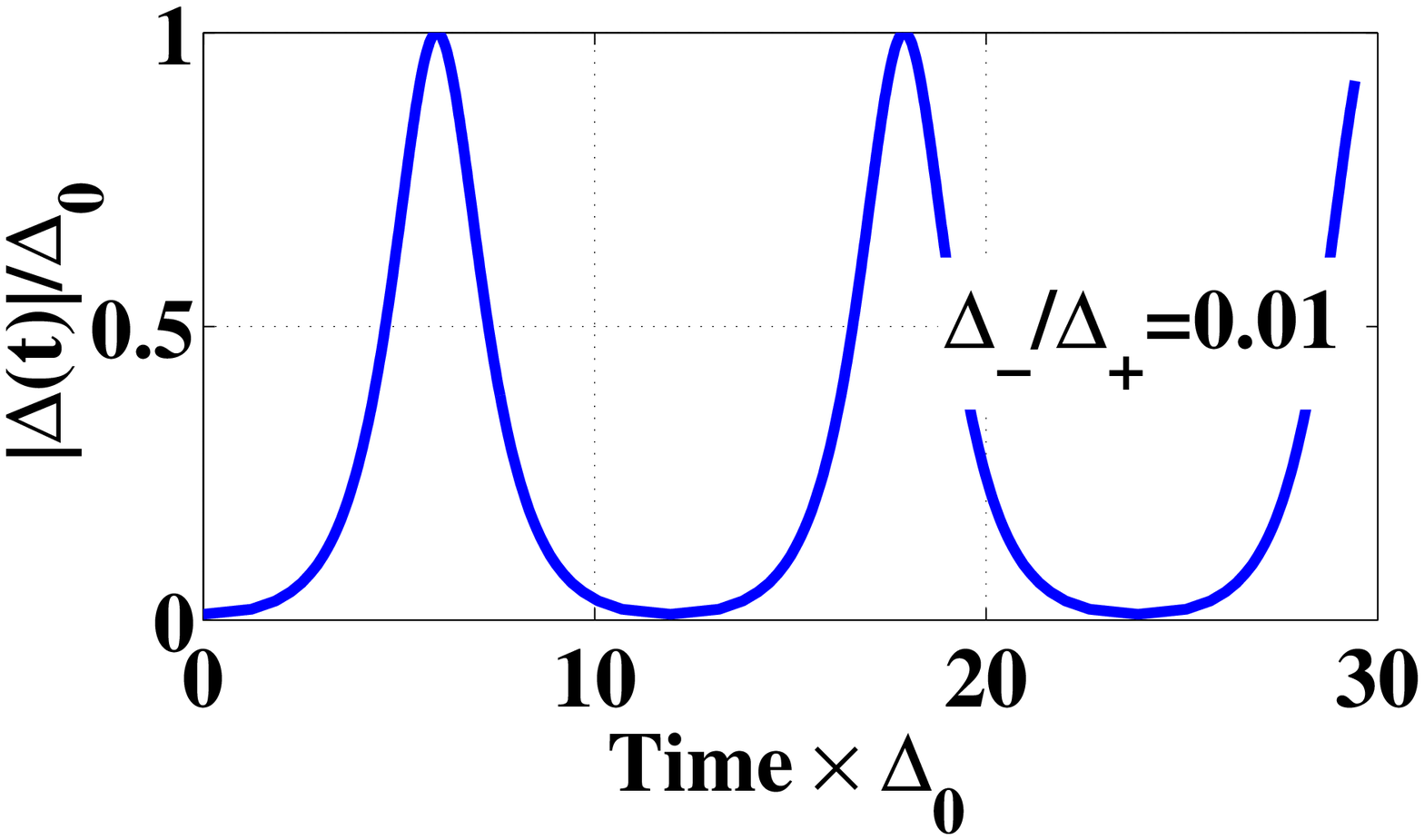}
\end{minipage}
\hspace{-1.3in}
\begin{minipage}[t]{1.2in}
\vspace{0.75in}
\includegraphics[width=1.2in]{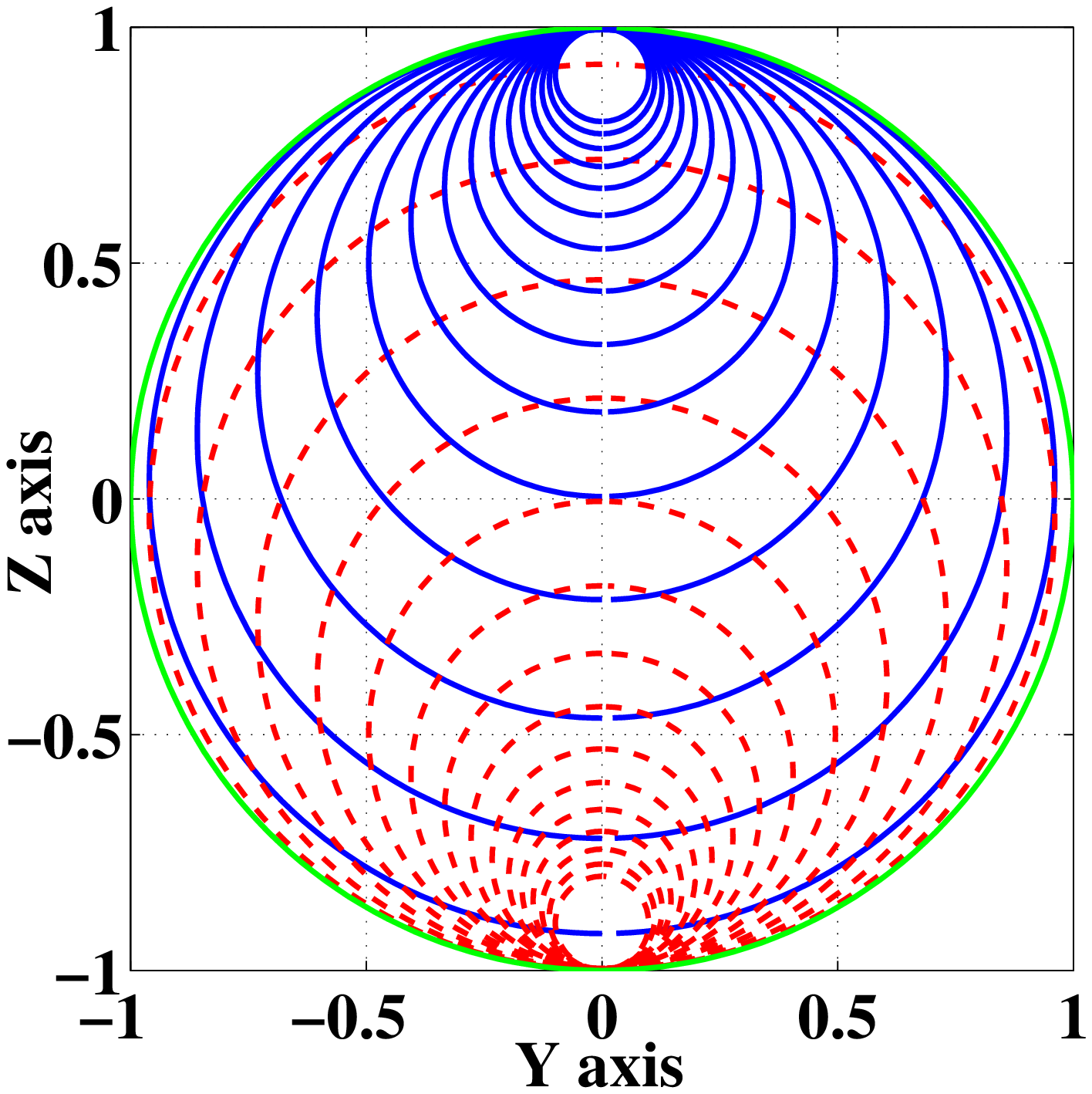}
\end{minipage}
\begin{minipage}[t]{1.2in}
\vspace{0.0in}
\centering
\includegraphics[width=1.2in]{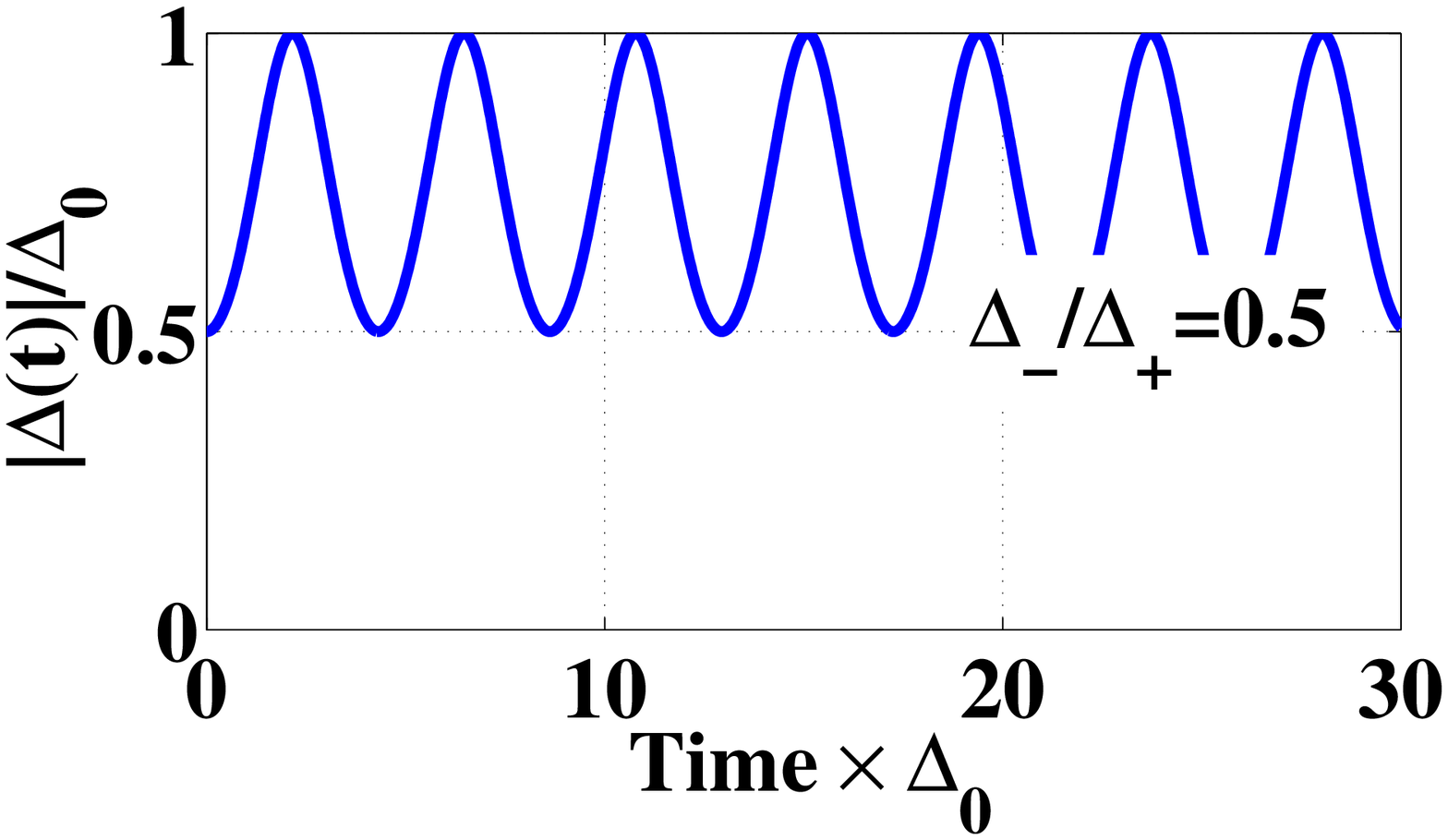}
\end{minipage}
\hspace{-1.3in}
\begin{minipage}[t]{1.2in}
\vspace{0.75in}
\includegraphics[width=1.2in]{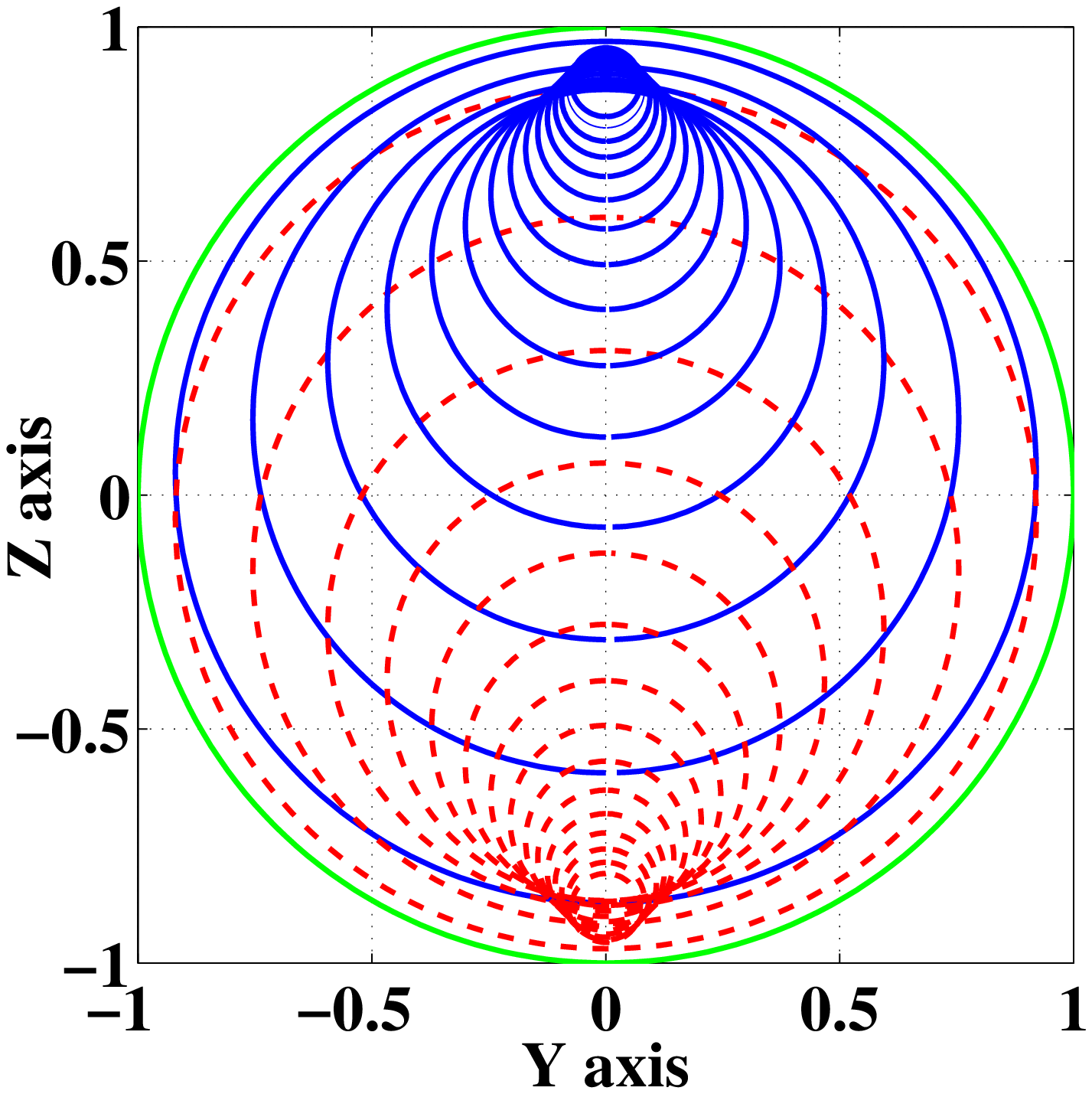}
\end{minipage}
\begin{minipage}[t]{1.2in}
\vspace{0.0pt}
\centering
\includegraphics[width=1.2in]{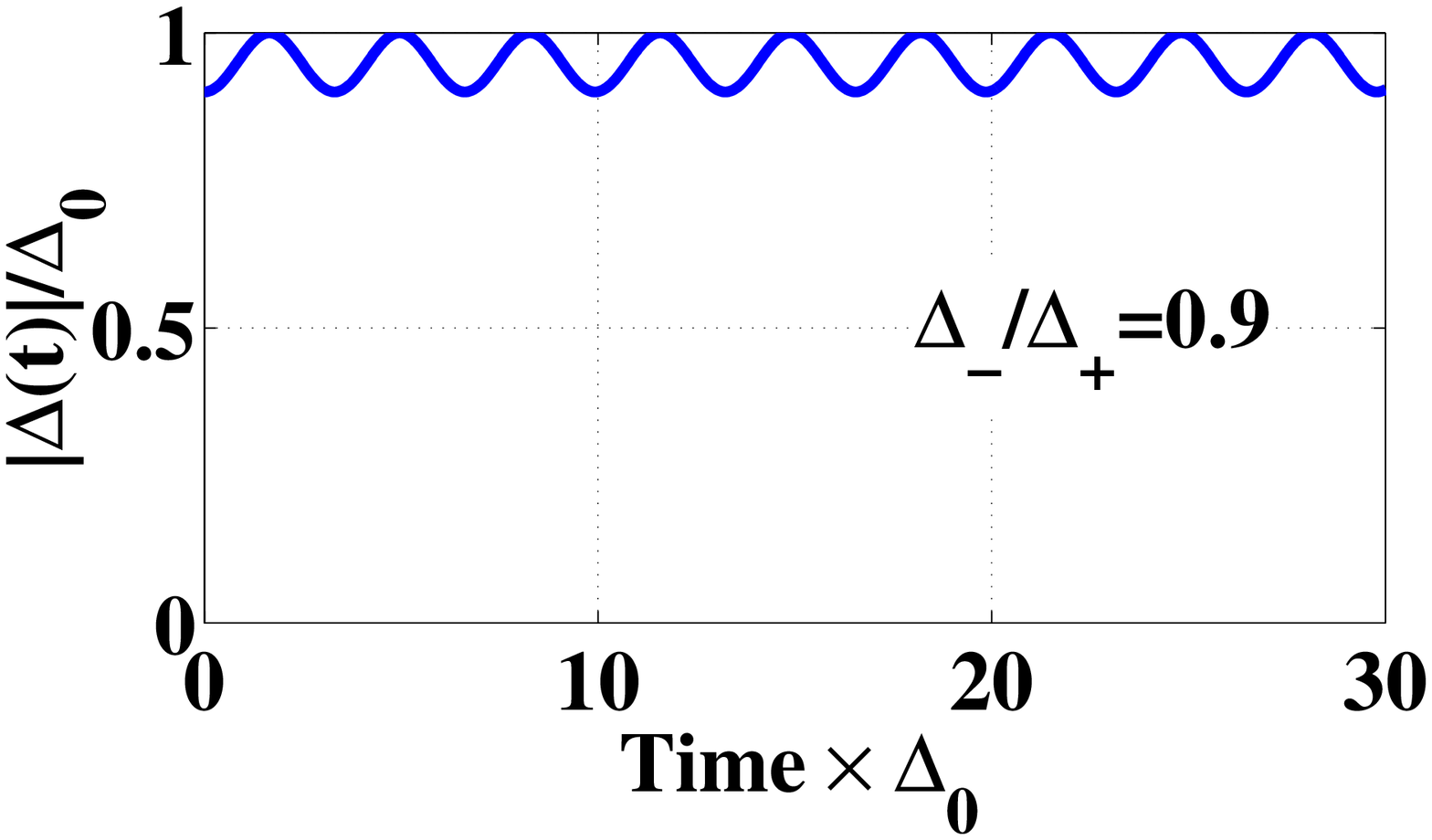}
\end{minipage}
\hspace{-1.3in}
\begin{minipage}[t]{1.2in}
\vspace{0.75in}
\includegraphics[width=1.2in]{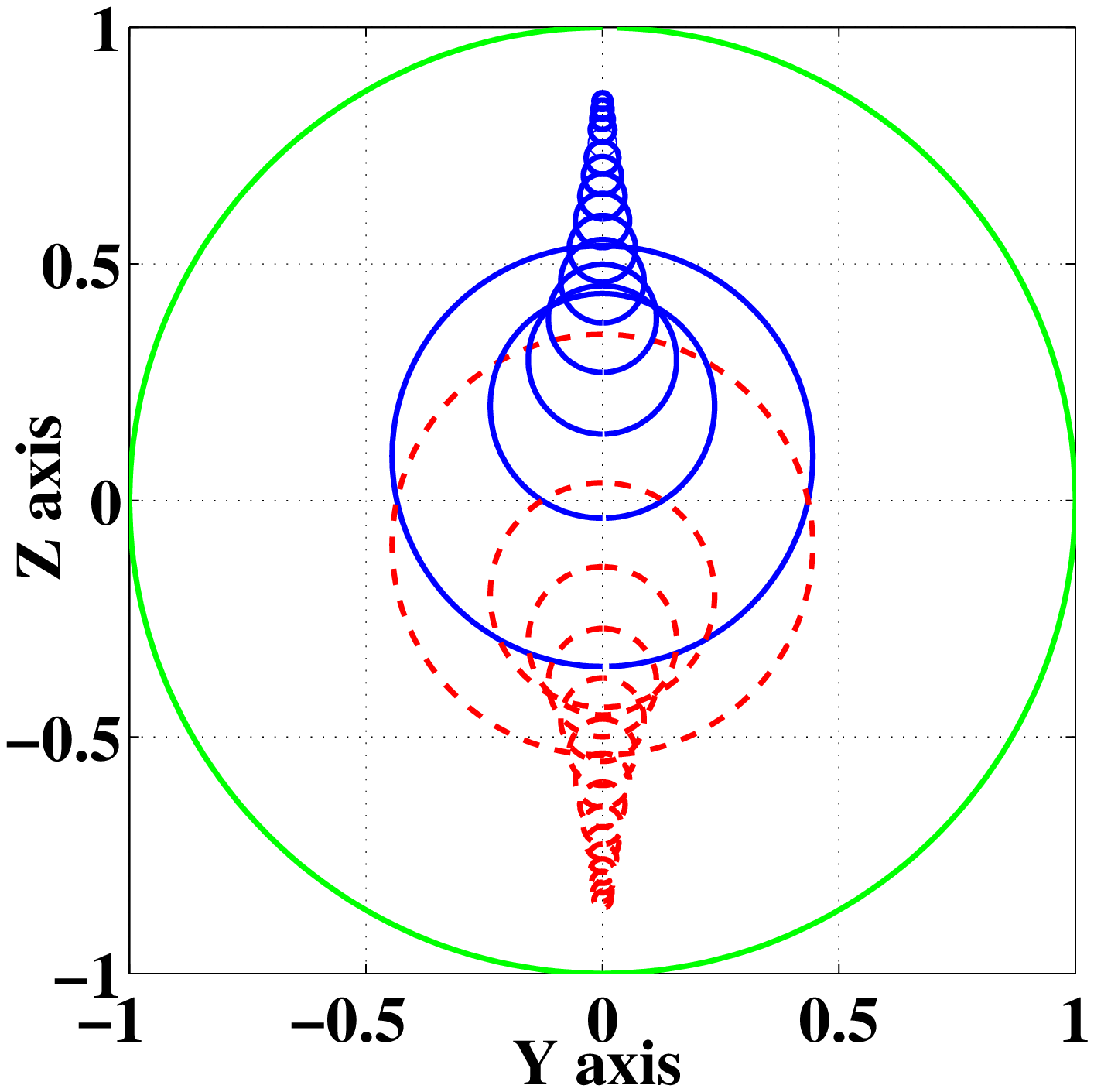}
\end{minipage}
}
\vspace{0.15cm}
\caption[]{
Coherent BCS dynamics on the Bloch sphere (\ref{eq:Bloch_sphere}).
Trajectories of individual Cooper pair states
for the soliton train solutions
(Eq.\,(\ref{eq:Omega_Delta_01}) and insets)
oscillating in the limits
$\Delta_-\le\Delta(t)\le\Delta_+$
are shown.
Note that each state completes a full
$2\pi$
Rabi cycle per soliton.
The red and blue curves correspond to the energies $\epsilon_{\vec p}$
above and below the Fermi level.
}
\label{fig:Bloch_sphere}
\end{figure}

Motivated by the stability analysis (\ref{eq:exp(gamma_t)}),
(\ref{eq:instability_exponent}),
we try the ansatz
$\Delta(t)=e^{-i\omega t}\alpha^{-1}(t)$ with $\alpha(t)$ real,
and
\be\label{eq:w_p_ansatz}
w_{\vec p}(t)=2\epsilon_{\vec p}f(t)-i\dot f(t)
,\quad
f(t)\equiv \frac1{\Delta^\ast}
=e^{-i\omega t}\alpha(t)
\ee
Substituting this in Eq.(\ref{eq:w_p}), we obtain an equation
\be\label{eq:alpha_long}
i\xi_{\vec p}\dot\alpha+\ddot\alpha
= \xi_{\vec p} (\xi_{\vec p}\alpha-i\dot\alpha)
+ \frac1{\alpha}\lp 1 - (\xi_{\vec p}\alpha-i\dot\alpha)^2\rp
\ee
with $\xi_{\vec p}=2\epsilon_{\vec p}-\omega$.
Remarkably,
the terms with $\xi_{\vec p}$ cancel, and
Eq.\,(\ref{eq:alpha_long}) takes the same form for all the states,
\be\label{eq:alpha}
\alpha \ddot\alpha = \dot\alpha^2+1
\ee
which justifies the ansatz (\ref{eq:w_p_ansatz}).
By a variable substitution
$\alpha=e^\phi$,
Eq.\,(\ref{eq:alpha}) can be brought
to the form $\ddot\phi=e^{-2\phi}$.
Integrating the latter equation, 
obtain $\dot\phi^2+e^{-2\phi}=\gamma^2$,
with $\gamma$ an integration constant.
This yields $\dot\alpha^2=\gamma^2\alpha^2-1$,
\be\label{eq:cosh_soliton}
\alpha(t)=\frac1{\gamma}\cosh \gamma (t-t_{0})
,\quad
\Delta(t)
=\frac{\gamma e^{-i\omega t}}{\cosh \gamma (t-t_{0})}
\ee
The modulus $|\Delta|$ time dependence (Fig.\ref{fig:Bloch_sphere}, upper left),
growing first, then decreasing and taking the system back
to the unpaired state, reflects the absence of dissipation.
The time  $t_0$ at which $|\Delta|$ peaks is set by the initial condition
at large negative
times $t\sim t_\ast$.

For $\epsilon_{\vec p}<0$, $w_{\vec p}=v_{\vec p}/u_{\vec p}$,
the form of Eq.\,(\ref{eq:w_p}) remains the same up to a sign change
$\epsilon_{\vec p}\to -\epsilon_{\vec p}$ and the permutation
$\Delta\leftrightarrow\Delta^\ast$.
Accordingly, the ansatz for $w_{\vec p}$ in this case is
$w_{\epsilon_{\vec p}<0}(t)=2|\epsilon_{\vec p}|f(t)-i\dot f(t)$,
$f(t)\equiv\Delta^{-1}=e^{i\omega t}\alpha(t)$.
The resulting equation for $\alpha(t)$ is identical to
Eq.\,(\ref{eq:alpha}).

The last step is to analyze the requirements
on this solution
due to the selfconsistency condition
(\ref{eq:Delta_BCS}). For that, we rewrite Eq.\,(\ref{eq:Delta_BCS})
in terms of $w_{\vec p}(t)$ as
\be\label{eq:Delta_w}
\Delta(t)=
\lambda\sum_{\epsilon_{\vec p}>0}\frac{w_{\vec p}(t)}{1+|w_{\vec p}(t)|^2}
+
\lambda\sum_{\epsilon_{\vec p}<0}\frac{w_{\vec p}^\ast(t)}{1+|w_{\vec p}(t)|^2}
\ee
and note that both the right and the left hand side
have the same time dependence, and are equal to each other provided that
the quantity $\zeta = \omega +i\gamma$ satisfies
Eq.\,(\ref{eq:instability_exponent}). This means that 
the gap modulus $|\Delta(t)|$
peak value 
is equal to the instability exponent $\gamma$ defined
by (\ref{eq:instability_exponent}).

\vspace{-0.5cm}

At $T=0$, in the case of a constant density of states,
$\gamma$ is equal to the equilibrium BCS gap $\Delta_0$, while
$\omega$ vanishes
due to particle-hole symmetry.
Thus, remarkably, the modulus $|\Delta(t)|$ in this case
peaks exactly at $\Delta_0$.

To illustrate the collective dynamics in the soliton solution,
we plot the trajectories $u_{\vec p}(t)$, $v_{\vec p}(t)$ on the
Bloch sphere $r_1^2+r_2^2+r_3^2=1$ using the parameterization
\be\label{eq:Bloch_sphere}
r_1+ir_2=2u_{\vec p}v^\ast_{\vec p};\quad r_3=|u_{\vec p}|^2-|v_{\vec p}|^2
\ee
(Fig.\,\ref{fig:Bloch_sphere}, left).
Here, as well as in the soliton train solutions discussed below
(Eq.\,(\ref{eq:Omega_Delta_01})),
each state $(u_{\vec p},v_{\vec p})$ completes a full
Rabi cycle
per soliton. The trajectories, which are small loops for the pairs with
large energies $\epsilon_{\vec p}$, turn into
a big circle as
$\epsilon_{\vec p}$ tends to the Fermi level.

To gain further insight,
we reformulate the Bogoliubov approach, following Anderson \cite{Anderson58},
in terms of pseudospins associated with individual Cooper pair states.
`Pauli spin' operators
$\sigma^{\pm}_{\vec p}\equiv \frac12(\sigma^x_{\vec p}\pm i\sigma^y_{\vec p})$
can be assigned to each pair of
fermions with opposite momenta as follows
\be
\label{eq:pseudospins}
\sigma^+_{\vec p}
=a^+_{\vec p\uparrow}a^+_{-\vec p\downarrow}
\,,\quad
\sigma^-_{\vec p}
=a_{-\vec p\downarrow}a_{\vec p\uparrow}
\ee
The identity
$\sigma^z_{\vec p}\equiv [\sigma^+_{\vec p},\sigma^-_{\vec p}]
=a^+_{\vec p\uparrow}a_{\vec p\uparrow}
-a_{-\vec p\downarrow}a^+_{-\vec p\downarrow}$
allows to map the problem (\ref{eq:Hbcs}) onto an interacting spin problem
%
\be\label{eq:Hspin}
\HH={\sum_{\vec p}}'\epsilon_{\vec p}\sigma^z_{\vec p}
-2\lambda{\sum_{\vec p,\vec q}}'\sigma^+_{\vec p}\sigma^-_{\vec q}
\ee
where ${\sum}'_{\vec p}$ means a sum over the pairs of states
$(\vec p,-\vec p)$.
Since all the spins interact with each other equally,
the mean field theory here is exact, just like for the BCS problem.
The mean field hamiltonian for each spin is
\be\label{eq:Hp}
\HH_{\vec p}=\vec b_{\vec p}\cdot\vec{\s_{\vec p}}
,\quad
\vec b_{\vec p}=
\lp -\Delta', - \Delta'', \epsilon_{\vec p}\rp
\ee
Here the $z$ component of the effective field $\vec b_{\vec p}$,
given by the single particle energy,
is spin-specific, while the transverse components,
the same for all of the spins, satisfy
\be\label{eq:Delta_sigma_p}
\Delta\equiv \Delta'+i\Delta'' =
\lambda \sum_{\vec q}  \la  \s^+ _{\vec q}\ra
\ee
which is analogous to the BCS gap relation.
In the ground state each spin is aligned with $\vec b_{\vec p}$,
and the spins form a texture near the Fermi surface
\cite{Anderson58},
with spin rotation described by the Bogoliubov angle.

The dynamical problem of interest can be cast in the form
of Bloch equations for the spins,
\be\label{eq:Bloch_sigma}
\dot \s_{\vec p} =
i[\HH_{\vec p},\s_{\vec p}]=2 \vec b_{\vec p} \times \vec \s_{\vec p}
\ee
with the field $\vec b_{\vec p}$ defined selfconsistently by
(\ref{eq:Hp}),(\ref{eq:Delta_sigma_p}).
Anderson~\cite{Anderson58} used Eq.\,(\ref{eq:Bloch_sigma}),
linearized about the texture
ground state,
to study collective excitations of
a superconductor (see also~\cite{Volkov74}). Linearized about the unpaired
state, Eq.\,(\ref{eq:Bloch_sigma}) 
yields an instability identical to (\ref{eq:exp(gamma_t)}),\,(\ref{eq:instability_exponent}).

\begin{figure}[t]
\centerline{
\begin{minipage}[t]{3.5in}
\vspace{0.2pt}
\centering
\includegraphics[width=3.5in]{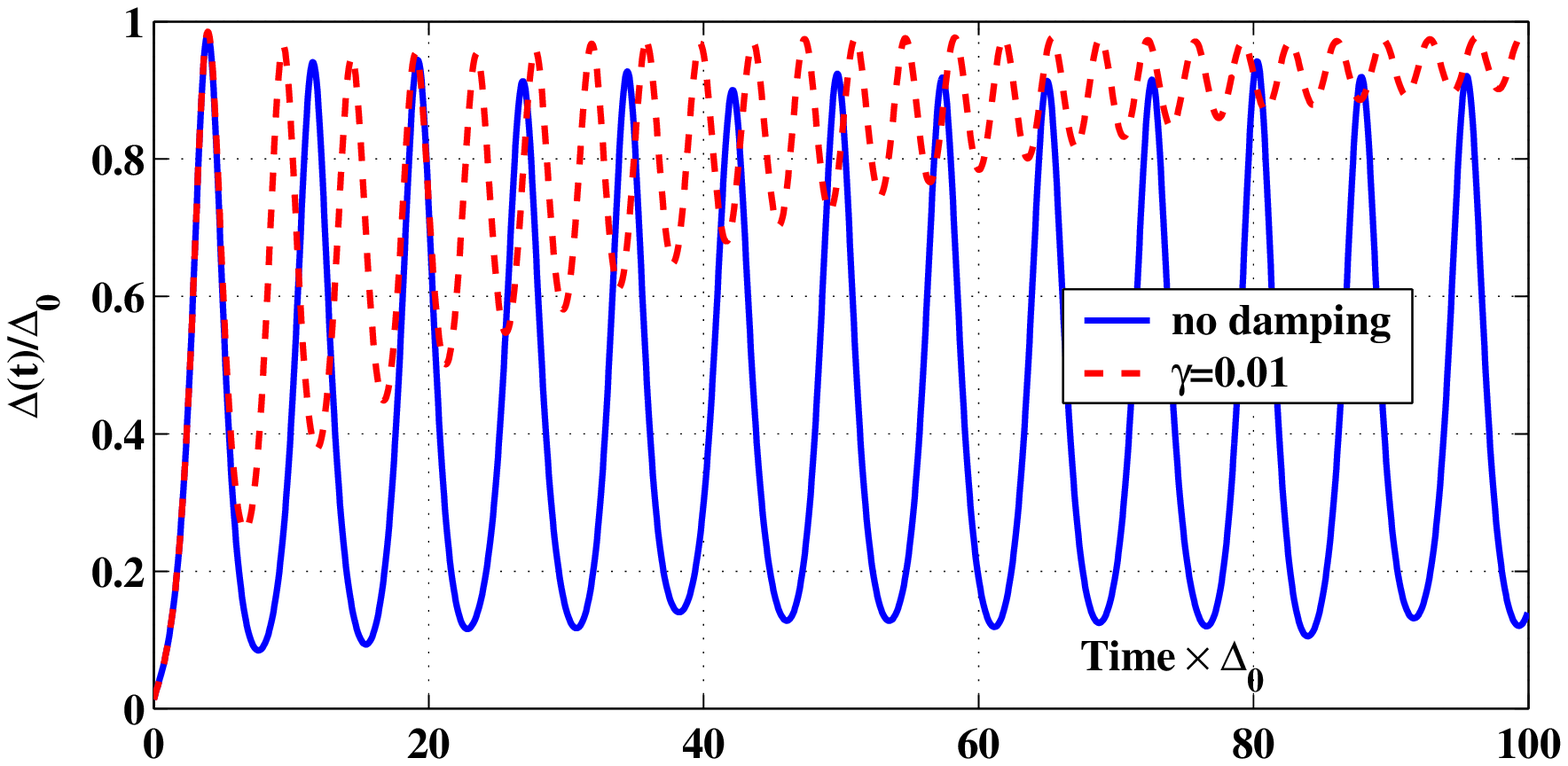}
\end{minipage}
}
\vspace{0cm}
\caption[]{
Bloch dynamics $\dot \vec r_{\vec p}=2\vec b_{\vec p}^{\rm eff}\times \vec r_{\vec p}$
for $10^3$ spins with a constant density of states,
with $\vec b_{\vec p}^{\rm eff}$ defined by (\ref{eq:Hp}),(\ref{eq:b_gamma}). 
Initial conditions
$r_{3,\vec p}=\tanh(\textstyle{\frac12}\beta\epsilon_{\vec p})$,
$(r_{1}\!+\!i r_{2})_{\vec p}=e^{i\phi_{\vec p}}(1-r_{3}^2)^{1/2}$,
with $\beta^{-1}\equiv T=0.1\Delta_0$ 
and random uncorrelated $0<\phi_{\vec p}<2\pi$
were used to simulate thermal noise. 
}
\label{fig:simulation}
\end{figure}

The Bloch dynamics is suitable for simulation
(Fig.\,\ref{fig:simulation}),
since Eq.\,(\ref{eq:Bloch_sigma}) is
linear in spin operators and can
be replaced by an equation
for the expectation values (\ref{eq:Bloch_sphere}).
Typical $\Delta(t)$, observed in the presence of thermal noise,
is an orderly sequence of
the $\cosh{}$ solitons (Fig.\,\ref{fig:simulation}, top),
indicating robustness of soliton solution
({\it cf.} Ref.\cite{Galperin81}).

\vspace{-0.5cm}

The effect of damping can be studied by replacing
\be\label{eq:b_gamma}
\vec b_{\vec p}
\to
\vec b_{\vec p}^{\rm eff}=\vec b_{\vec p}-\gamma \vec b_{\vec p}\times\vec r_{\vec p}
\ee
with a dimensionless damping constant,
$\gamma\sim 1/(\Delta_0\tau_{\epsilon})$.
This destroys integrability and prompts relaxation to BCS equilibrium 
(Fig.\,\ref{fig:simulation}, bottom). 

In our nonlinear dynamical problem,
we have to solve the Bloch equations (\ref{eq:Bloch_sigma})
together with the selfconsistency
relation (\ref{eq:Delta_sigma_p}).
As above, we assume the pairing amplitude
time dependence of the form $\Delta(t)=e^{-i\omega t}\Omega(t)$, where
$\Omega(t)$ is real and
$\omega$ is a constant frequency shift to be fixed by the
relation (\ref{eq:Delta_sigma_p}).
In the Larmor frame rotating with the frequency $\omega$,
Eq.\,(\ref{eq:Bloch_sigma}), written 
for the spins average polarization $r_i=\la \s^i_{\vec p}\ra$ components, reads
\be\label{eq:r_123}
\dot r_1=-\xi_{\vec p}r_2
,\quad
\dot r_2=\xi_{\vec p}r_1+2\Omega r_3
,\quad
\dot r_3=-2\Omega r_2
\ee
with $\xi_{\vec p}=2\epsilon_{\vec p}-\omega$, as before.
An exact solution can be obtained from the ansatz
\be\label{eq:r_123_ansatz}
r_1=A\Omega
,\quad
r_2=B\dot\Omega
,\quad
r_3=C\Omega^2-D
\ee
The first and the third equation (\ref{eq:r_123}) are satisfied
by (\ref{eq:r_123_ansatz}) provided
$A=-\xi_{\vec p}B$ and $B=-C$,
while the second equation (\ref{eq:r_123})
is consistent with the normalization condition $r_1^2+r_2^2+r_3^2=1$,
and thus yields
\be\label{eq:Omega_normalization}
C^2\xi_{\vec p}^2\Omega^2 + C^2\dot\Omega^2 + (C\Omega^2-D)^2=1
\ee
Eq.\,(\ref{eq:Omega_normalization}) will take the same form for all the spins,
\be\label{eq:Omega_Delta_01}
\dot\Omega^2 +(\Omega^2-\Delta^2_-)(\Omega^2-\Delta^2_+)=0
,\quad
\Delta_- \le \Delta_+
\ee
provided that the constants $D$, $C$ are chosen as
\be\label{eq:CD_Delta_01}
(D^2-1)/C^2=\Delta^2_-\Delta^2_+
,\quad
2D/C=\xi_{\vec p}^2+\Delta^2_-+\Delta^2_+
\ee
with the sign factor ${\rm sgn}\,\epsilon_{\vec p}$.
Eq.\,(\ref{eq:Omega_Delta_01}) defines
an elliptic function $\Omega(t)$
oscillating periodically between
$\Delta_-$ and $\Delta_+$. At $\Delta_-\ll\Delta_+$,
the solution is a train
of weakly overlapping solitons 
(\ref{eq:cosh_soliton}) with $\Delta_+ =\gamma$
(Fig.\,\ref{fig:Bloch_sphere}, left).

The real 
part of the selfconsistency
relation (\ref{eq:Delta_sigma_p}),
\be\label{eq:Delta_01}
1=\lambda\sum_{\vec p}
\frac{\xi_{\vec p}{\rm sgn}\,\epsilon_{\vec p}}{\lp
(\xi_{\vec p}^2+\Delta^2_- +\Delta^2_+)^2-4\Delta^2_-\Delta^2_+\rp^{1/2}}
\ee
fixes one of the constants $\Delta_{\pm}$, the other one being a free
parameter that controls the inter-soliton
time separation.
With the ratio $\Delta_-/\Delta_+$ varying from $0$ to $1$,
the soliton frequency increases, and the weakly overlapping solitons
(\ref{eq:cosh_soliton}) gradually become overlapping more strongly, 
turning into weak harmonic oscillations
(Fig.\,\ref{fig:Bloch_sphere}).

We note that the weak oscillation limit has been investigated
by Volkov and Kogan~\cite{Volkov74} with the help of linearized Gorkov
equation, and earlier by Anderson~\cite{Anderson58}, who used
pseudospin representation. 
The nonexponential decay of the lianearized oscillations~\cite{Schmid66,Volkov74} 
was interpreted as collisionless damping, 
related to strong mixing of the
oscillations of $\Delta$ with the states of two excited quasiparticles 
slightly above the superconducting gap.

The imaginary part of Eq.\,(\ref{eq:Delta_sigma_p})
fixes the value of the frequency shift $\omega$ (we recall that
$\omega\ne0$ in the presence of charge asymmetry).
At $\Delta_-\ll\Delta_+$, Eq.\,(\ref{eq:Delta_01}) turns into
Eq.\,(\ref{eq:instability_exponent})
which, as we found above, defines the amplitude of a single soliton.
In the opposite limit,
$\Delta_-\to\Delta_+$, Eq.\,(\ref{eq:Delta_01})
coincides with the BCS gap equation.

There is an interesting relation between our problem
and the self-induced transparency
phenomenon~\cite{McCall67}.
In the latter, 
an optical pulse interacting with an ensemble of atoms
can dissipate its energy by inducing resonant Rabi
transitions in the atoms. However, when the pulse duration is tuned
so that the atoms complete a full Rabi $2\pi$ cycle as the pulse goes by,
the pulse sustains itself and propagates without dissipation. The Bloch equations
describing this phenomenon bear striking similarity 
to our Eqs.\,(\ref{eq:Bloch_sigma}),
while the atoms'
polarization is employed~\cite{McCall67} 
to provide feedback on the pulse instead of our BCS selfconsistency
relation.

Before concluding, we note that the dynamics
at finite temperature, in the regime described
by Eq.\,(\ref{eq:tau_epsilon>tau_delta}), remains
an open problem. In particular, we can not rule out the
possibility of chaotic behavior. 
At $T=0$ the problem has a relatively simple solution,
periodic in time, because in this case the quasiparticles with low energies
are strongly coupled to the oscillations of $\Delta$. In
contrast, in the case $T\tau_\Delta\gg 1$ there are two groups of
quasiparticles: those with energies $\epsilon_{\vec p}\sim \tau_\Delta^{-1}$,
which fully participate in
the oscillations, as above,
and the quasiparticles  with  $\tau_\Delta^{-1} \ll \epsilon_{\vec p}\ll T$,
coupled to $\Delta(t)$ much weaker and playing 
the role of a thermal bath, thereby providing dissipation.

In summary, this work provides an exact solution for
the BCS pairing formation problem. 
In the nonadiabatic regime, the dynamics is dissipationless and thus nonlinear.
Soliton train solutions are obtained analytically and demonstrated 
to be generic and robust by a simulation.


\end{multicols}

\end{document}